\documentclass[prl,reprint,superscriptaddress]{revtex4-1}

\usepackage{color}
\usepackage{graphicx}
\usepackage{amsmath}
\usepackage{array}
\begin{document}

\title{Constraining Dark Matter Models with a Light Mediator from PandaX-II Experiment}
\date{\today}
\author{Xiangxiang Ren}
\affiliation{INPAC and School of Physics and Astronomy, Shanghai Jiao Tong University, Shanghai Laboratory for Particle Physics and Cosmology, Shanghai 200240, China}
\author{Li Zhao}
\affiliation{INPAC and School of Physics and Astronomy, Shanghai Jiao Tong University, Shanghai Laboratory for Particle Physics and Cosmology, Shanghai 200240, China}
\author{Abdusalam Abdukerim}
\affiliation{School of Physics and Technology, Xinjiang University, \"{U}r\"{u}mqi 830046, China}
\author{Xun Chen}
\affiliation{INPAC and School of Physics and Astronomy, Shanghai Jiao Tong University, Shanghai Laboratory for Particle Physics and Cosmology, Shanghai 200240, China}
\author{Yunhua Chen}
\affiliation{Yalong River Hydropower Development Company, Ltd., 288 Shuanglin Road, Chengdu 610051, China}
\author{Xiangyi Cui}
\affiliation{INPAC and School of Physics and Astronomy, Shanghai Jiao Tong University, Shanghai Laboratory for Particle Physics and Cosmology, Shanghai 200240, China}
\author{Deqing Fang}
\affiliation{Shanghai Institute of Applied Physics, Chinese Academy of Sciences, Shanghai 201800, China}
\author{Changbo Fu}
\author{Karl Giboni}
\affiliation{INPAC and School of Physics and Astronomy, Shanghai Jiao Tong University, Shanghai Laboratory for Particle Physics and Cosmology, Shanghai 200240, China}
\author{Franco Giuliani}
\author{Linhui Gu}
\affiliation{INPAC and School of Physics and Astronomy, Shanghai Jiao Tong University, Shanghai Laboratory for Particle Physics and Cosmology, Shanghai 200240, China}
\author{Xuyuan Guo}
\affiliation{Yalong River Hydropower Development Company, Ltd., 288 Shuanglin Road, Chengdu 610051, China}
%\author{Zhifan Guo}
%\affiliation{School of Mechanical Engineering, Shanghai Jiao Tong University, Shanghai 200240, China}
\author{Ke Han}
\author{Changda He}
\author{Di Huang}
\affiliation{INPAC and School of Physics and Astronomy, Shanghai Jiao Tong University, Shanghai Laboratory for Particle Physics and Cosmology, Shanghai 200240, China}
\author{Shengming He}
\affiliation{Yalong River Hydropower Development Company, Ltd., 288 Shuanglin Road, Chengdu 610051, China}
\author{Xingtao Huang}
\affiliation{School of Physics and Key Laboratory of Particle Physics and Particle Irradiation (MOE), Shandong University, Jinan 250100, China}
\author{Zhou Huang}
\affiliation{INPAC and School of Physics and Astronomy, Shanghai Jiao Tong University, Shanghai Laboratory for Particle Physics and Cosmology, Shanghai 200240, China}
\author{Xiangdong Ji}
\email[Spokesperson: ]{xdji@sjtu.edu.cn}
\affiliation{INPAC and School of Physics and Astronomy, Shanghai Jiao Tong University, Shanghai Laboratory for Particle Physics and Cosmology, Shanghai 200240, China}
\affiliation{Department of Physics, University of Maryland, College Park, Maryland 20742, USA}
\affiliation{Tsung-Dao Lee Institute, Shanghai 200240, China}
\author{Yonglin Ju}
\affiliation{School of Mechanical Engineering, Shanghai Jiao Tong University, Shanghai 200240, China}
\author{Yao Li}
\author{Heng Lin}
\affiliation{INPAC and School of Physics and Astronomy, Shanghai Jiao Tong University, Shanghai Laboratory for Particle Physics and Cosmology, Shanghai 200240, China}
\author{Huaxuan Liu}
\affiliation{School of Mechanical Engineering, Shanghai Jiao Tong University, Shanghai 200240, China}
\author{Jianglai Liu}
\affiliation{INPAC and School of Physics and Astronomy, Shanghai Jiao Tong University, Shanghai Laboratory for Particle Physics and Cosmology, Shanghai 200240, China}
\affiliation{Tsung-Dao Lee Institute, Shanghai 200240, China}
\author{Yugang Ma}
\affiliation{Shanghai Institute of Applied Physics, Chinese Academy of Sciences, Shanghai 201800, China}
\author{Yajun Mao}
\affiliation{School of Physics, Peking University, Beijing 100871, China}
\author{Kaixiang Ni}
\affiliation{INPAC and School of Physics and Astronomy, Shanghai Jiao Tong University, Shanghai Laboratory for Particle Physics and Cosmology, Shanghai 200240, China}
\author{Jinhua Ning}
\affiliation{Yalong River Hydropower Development Company, Ltd., 288 Shuanglin Road, Chengdu 610051, China}
\author{Andi Tan}
\affiliation{Department of Physics, University of Maryland, College Park, Maryland 20742, USA}
\author{Hongwei Wang}
\affiliation{Shanghai Institute of Applied Physics, Chinese Academy of Sciences, Shanghai 201800, China}
\author{Meng Wang}
\affiliation{School of Physics and Key Laboratory of Particle Physics and Particle Irradiation (MOE), Shandong University, Jinan 250100, China}
\author{Qiuhong Wang}
\affiliation{Shanghai Institute of Applied Physics, Chinese Academy of Sciences, Shanghai 201800, China}
\author{Siguang Wang}
\affiliation{School of Physics, Peking University, Beijing 100871, China}
\author{Xiuli Wang}
\affiliation{School of Mechanical Engineering, Shanghai Jiao Tong University, Shanghai 200240, China}
\author{Shiyong Wu}
\affiliation{Yalong River Hydropower Development Company, Ltd., 288 Shuanglin Road, Chengdu 610051, China}
\author{Mengjiao Xiao}
\affiliation{Department of Physics, University of Maryland, College Park, Maryland 20742, USA}
\affiliation{Center of High Energy Physics, Peking University, Beijing 100871, China}
\author{Pengwei Xie}
\affiliation{INPAC and School of Physics and Astronomy, Shanghai Jiao Tong University, Shanghai Laboratory for Particle Physics and Cosmology, Shanghai 200240, China}
\author{Binbin Yan}
\affiliation{School of Physics and Key Laboratory of Particle Physics and Particle Irradiation (MOE), Shandong University, Jinan 250100, China}
\author{Jijun Yang}
\author{Yong Yang}
\email[Corresponding author: ]{yong.yang@sjtu.edu.cn}
\affiliation{INPAC and School of Physics and Astronomy, Shanghai Jiao Tong University, Shanghai Laboratory for Particle Physics and Cosmology, Shanghai 200240, China}
\author{Hai-Bo Yu}
\affiliation{Department of Physics and Astronomy, University of California, Riverside, California 92521, USA}
\author{Jianfeng Yue}
\affiliation{Yalong River Hydropower Development Company, Ltd., 288 Shuanglin Road, Chengdu 610051, China}
\author{Tao Zhang}
\affiliation{INPAC and School of Physics and Astronomy, Shanghai Jiao Tong University, Shanghai Laboratory for Particle Physics and Cosmology, Shanghai 200240, China}
\author{Jifang Zhou}
\affiliation{Yalong River Hydropower Development Company, Ltd., 288 Shuanglin Road, Chengdu 610051, China}
\author{Ning Zhou}
\affiliation{INPAC and School of Physics and Astronomy, Shanghai Jiao Tong University, Shanghai Laboratory for Particle Physics and Cosmology, Shanghai 200240, China}
\author{Qibin Zheng}
\affiliation{School of Medical Instrument and Food Engineering, University of Shanghai for Science and Technology, Shanghai 200093, China}
\author{Xiaopeng Zhou}
\affiliation{School of Physics, Peking University, Beijing 100871, China}
\collaboration{PandaX-II Collaboration}
\begin{abstract}

We search for nuclear recoil signals of dark matter models with a
light mediator in PandaX-II, a direct detection experiment in China
Jinping underground Laboratory. Using data collected in 2016 and 2017
runs, corresponding to a total exposure of 54 ton day, we set upper
limits on the zero-momentum dark matter-nucleon cross section.  These
limits have a strong dependence on the mediator mass when it is
comparable to or below the typical momentum transfer. We apply our
results to constrain self-interacting dark matter models with a light
mediator mixing with standard model particles, and set strong limits
on the model parameter space for the dark matter mass ranging from
$5~{\rm GeV}$ to $10~{\rm TeV}$.

\end{abstract}
\maketitle

The existence of dark matter (DM) is supported by a wide range of
observations in astronomy and cosmology, but its particle nature
remains elusive. Leading candidates such as weakly interacting massive
particles (WIMPs)~\cite{Bertone:2004pz} that could explain the
observed cosmological DM abundance, have been actively searched for in
indirect and direct detection experiments, as well as at the Large
Hadron Collider. The direct WIMP searches often assume a point-like
contact interaction between the DM candidate and the nucleus, since
the momentum transfer in nuclear recoils is much smaller than the
weak-scale mediator mass. However, this assumption breaks down if the
DM-nucleus interaction is mediated by a force carrier that has a mass
comparable to or lighter than the momentum
transfer~\cite{Foot:2003iv,Fornengo:2011sz,
  Kaplinghat:2013yxa,Li:2014vza,DelNobile:2015uua,Kahlhoefer:2017ddj}.

Dark matter models with a light mediator are well-motivated. For
example, in many hidden-sector DM
models~\cite{Boehm:2003hm,Pospelov:2007mp,Feng:2008mu,Foot:2014uba,Blennow:2016gde}, DM particles
annihilate to the light mediator to achieve the observed abundance. It
can induce an attractive potential between two DM particles and boost
the annihilation cross
section~\cite{ArkaniHamed:2008qn,Pospelov:2008jd}. Furthermore, it has
been shown the self-interacting DM (SIDM) model with a light mediator
can explain observed stellar kinematics from dwarf galaxies to galaxy
clusters~\cite{Kamada:2016euw,Kaplinghat:2015aga}, a challenge for the
prevailing cold DM model (see, e.g.,~\cite{Tulin:2017ara}). If it
couples to the standard model (SM) particles, the DM signal event in direct
detection can be enhanced towards low recoil energies, a smoking-gun
signature of
SIDM~\cite{Kaplinghat:2013yxa,DelNobile:2015uua,Kahlhoefer:2017ddj,Baldes:2017gzu}. In
recent years, there has been great progress in the search for the
light force mediator at the high-luminosity facilities (see,
e.g.,~\cite{Battaglieri:2017aum} for a review).

In this {\it Letter}, we report upper limits on the DM-nucleon
scattering cross section induced by a light mediator and then
interpret them to constrain SIDM models proposed
in~\cite{Kaplinghat:2013yxa}. Our analysis is based on the data from
the PandaX-II experiment, which is the phase-II experiment in the
PandaX project that consists of a series of xenon-based rare-event
detection experiments, located at China Jinping underground Laboratory
(CJPL). The central apparatus of PandaX-II is a dual-phase xenon time
projection chamber (TPC). The active volume contains 580 kg liquid
xenon. Particle interacting with xenon results in prompt scintillation
photons ($S1$ signal) in liquid xenon as well as delayed
electroluminescence photons ($S2$ signal) in gaseous xenon. Both signals
are detected in one event by two arrays of photomultiplier tubes
(PMTs), located in the top and bottom of the TPC. More detailed
descriptions of the PandaX-II experiment can be found in
~\cite{Tan:2016diz,Tan:2016zwf,Cui:2017nnn}.

We first consider a general case, where DM interacts with the nucleon
through a vector or scalar force mediator, $\phi$, and further assume
$\phi$ has equal effective couplings to the proton and neutron as in
the standard WIMP model. The general form of the DM-nucleus elastic
scattering cross section can be parametrized
as~\cite{Kaplinghat:2013xca}

\begin{equation}
\label{eq:general}
\sigma(q^2)_{\chi N} =
\sigma|_{q^2=0}A^{2}\left(\frac{\mu}{\mu_{p}}\right)^{2}\frac{m^{4}_{\phi}}{(m^{2}_{\phi}+q^{2})^2}F^{2}(q^2),
\end{equation}
where $\sigma|_{q^2=0}$ is the DM-nucleon cross section in the limit
of zero momentum transfer ($q^2=0$), $A$ the mass number of the
nucleus, $\mu$ ($\mu_{p}$) the DM-nucleus (nucleon) reduced mass,
$m_\phi$ the mediator mass, and $F(q^2)$ the nuclear form factor. We
see that $\sigma_{\chi N}$ is momentum-dependent and it approaches the
standard WIMP case when $m_\phi\gg q$.

The differential recoil rate (in unit of counts per day per kg per
keV) is~\cite{Savage:2008er}
\begin{equation}
\frac{dR}{dE} = \frac{\sigma(q^2)_{\chi N}\rho}{2m_{\chi}\mu^{2}}\int_{v\geq  v_{\mathrm{min}}}d^{3}v v f(v,t)
  \label{eq:drde}
\end{equation}
where $\rho$ is the local DM density which we set to be 0.3 GeV/cm$^{3}$, $m_{\chi}$ is the DM
particle mass, $f(v,t)$ is the time-dependent DM velocity distribution
relative to the detector, and $v_{\mathrm{min}}$ is the minimum DM
velocity that results in a recoil energy $E$.

This analysis uses the same data sets as the recent WIMP search
(unblind) in PandaX-II~\cite{Cui:2017nnn}, consisting of 80 live day
of exposure in 2016 and 77 live day of exposure in 2017, the largest
published data set of its kind to date. We apply the same event
selection criteria as in Ref.~\cite{Cui:2017nnn}. The range for $S1$
and $S2$ signals are between 3 photoelectron (PE) and 45 PE, and 100
PE (uncorrected) and 10000 PE, respectively. The total data were
divided into 18 sets to take into account variations of detector
parameters and background rates. Background contributions have been
estimated, and no excess of events in data was observed above the
background. For a given DM model, the expected event distributions are
modeled with the same procedure as in~\cite{Cui:2017nnn}. For each
data set, we simulate the expected $S1$ and $S2$ signal distributions
from the SIDM recoil-energy spectra using a tuned NEST simulation
framework. Then, we apply the experimental efficiencies to make
further corrections.

\begin{figure}[!htbp]\centering
  \includegraphics[width=0.95\linewidth]{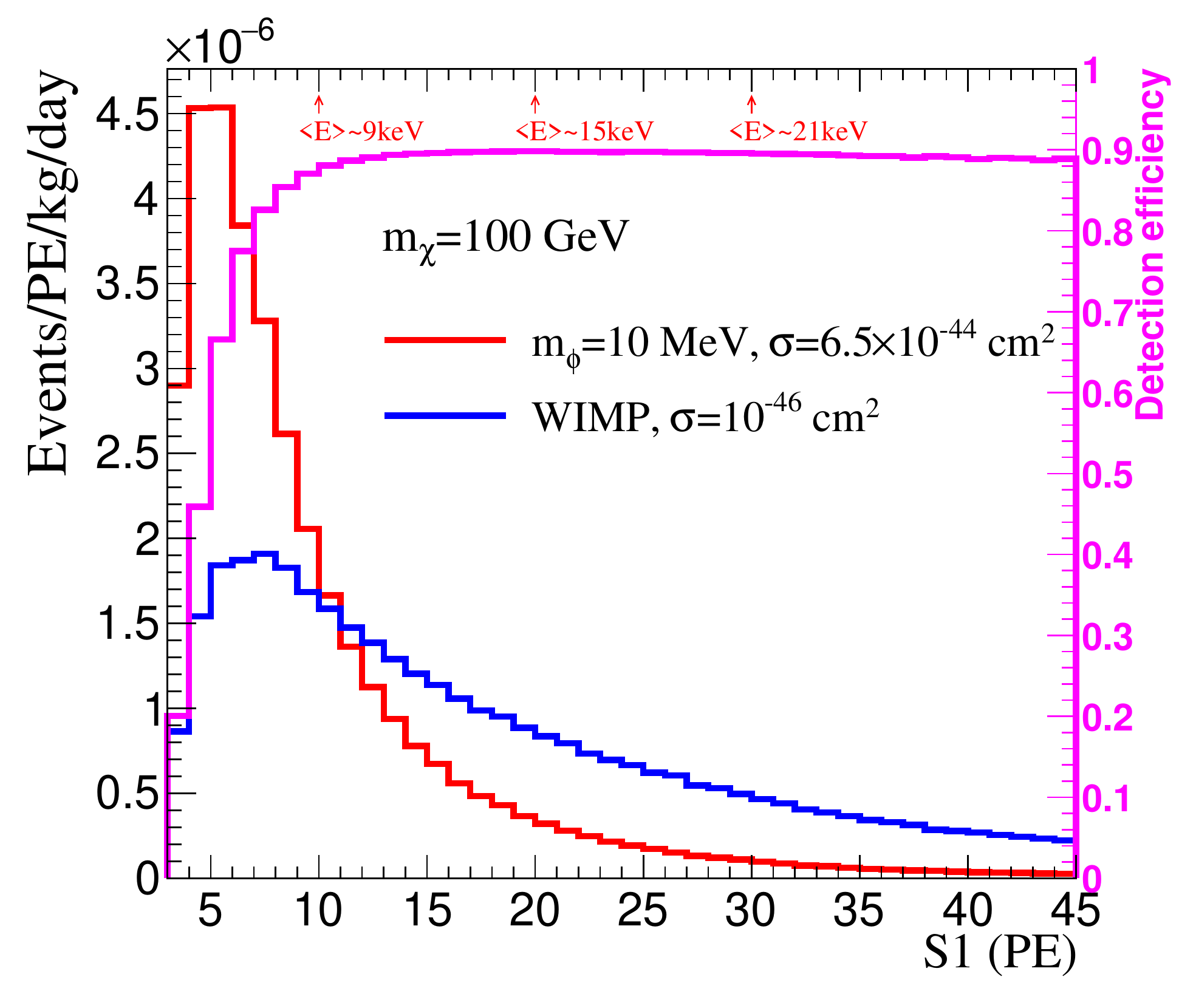}
  \caption{Simulated $S1$ spectrum (red) in PandaX-II for
    $(m_\chi,~m_\phi)=(100,~0.01)~{\rm GeV}$ with the zero-momentum DM-nucleon
    cross section fixed to be $6.5\times10^{-44}~{\rm cm^{2}}$. The distribution (blue) for the same DM mass but with a
    contact interaction is also shown. We take the WIMP-nucleon cross section to be  
    $10^{-46}~{\rm cm^{2}}$ so both spectra have the same integrated
    rate. The detection efficiency vs. $S1$ (magenta) is plotted with the corresponding value
    labeled on the right vertical axis. The rough conversion
    between $S1$ and recoil energy is indicated near the top
    horizontal axis for $S1$$=10,~20$ and $30$ PE.}
    \label{fig:s1}
\end{figure}

\begin{figure*}[!htbp]\centering
  \includegraphics[width=0.45\linewidth]{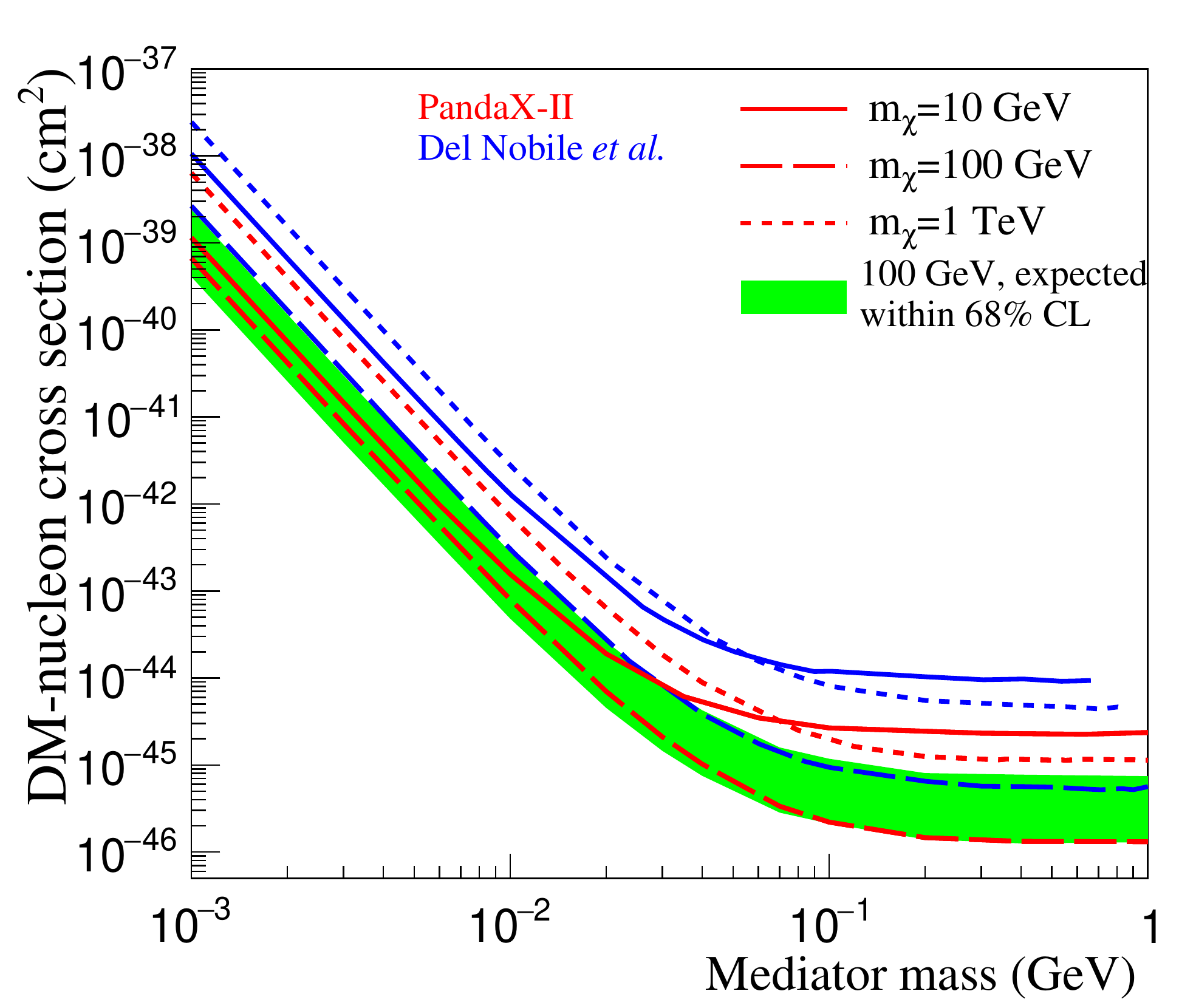}
  \includegraphics[width=0.45\linewidth]{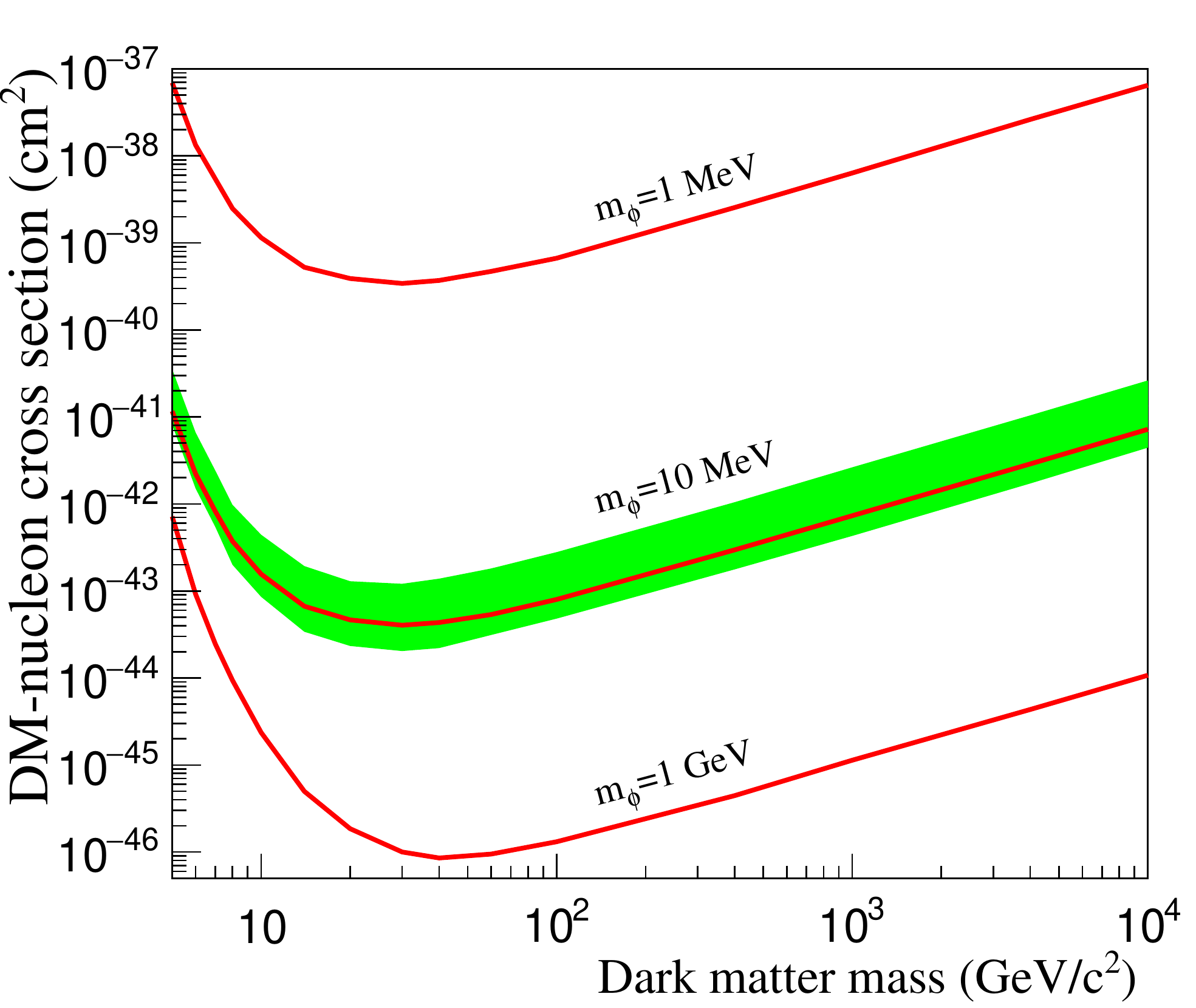}
  \caption{PandaX-II 90\% CL upper limits (red) on the zero-momentum
    DM-nucleon cross section for DM models with a light
    mediator. Left: the cross section limits vs. the mediator mass for
    three representative DM masses $10,~100~{\rm GeV}$ and $1~{\rm
      TeV}$. The limits converted from previous results
    from~\cite{DelNobile:2015uua} are included for comparison
    (blue). Right: the limits vs. DM mass for mediator masses
    $1,\;10~{\rm MeV}$ and $1~{\rm GeV}$.  The green bands denote the
    $\pm 1\sigma$ sensitivity for the given model parameters.}
    \label{fig:lim_xsmphi}
\end{figure*}

Figure~\ref{fig:s1} shows the simulated $S1$ distributions in PandaX-II
for a $100~{\rm GeV}$ DM particle with $m_\phi=$10 ${\rm MeV}$ (red)
and a WIMP with the same mass (blue). Both cases have the same
integrated rate, but their spectra are very different, i.e., the
$S1$ distribution of the light-mediator model is more peaked towards to
small $S1$ than predicted in the WIMP model. We also plot PandaX-II
detection efficiency as a function of $S1$ (magenta). It is nearly a
constant over the range of $10\textup{--}45$ PE, but is reduced
dramatically for $S1$ $<10$ PE, where the event rate of the
light-mediator model is maximized. Thus, we expect DM direct detection
sensitivity becomes weak when the mediator mass is comparable or less
than the typical momentum transfer in nuclear recoils, even though the
DM mass is still at the weak scale.

The same statistical method as in Ref~\cite{Cui:2017nnn} is used to
derive upper limits on signal cross section. An unbinned likelihood
function is constructed for these 18 data sets using the signal and
background probability density functions in the $S1$-$\log_{10}$($S2$/$S1$)
plane, taking into account the normalization uncertainties for signal
and background. For DM signal, we assign a conservative 20\%
uncertainty, estimated from different NEST simulations and
uncertainties on the detector parameters. The standard profile likelihood
ratio test statistic~\cite{Cowan:2010js,Aprile:2011hx} is evaluated at
grids of expected signal cross section (hypotheses) and compared to
the test statistic distribution obtained from large number of toy
Monte Carlo data produced and fitted using the same signal hypotheses.

Figure~\ref{fig:lim_xsmphi} shows the 90\% confidence level (CL) upper
limits on the zero-momentum DM-nucleon cross section assuming a light
mediator with mass ranging from $1~{\rm MeV}$ to $1~{\rm GeV}$. In the
left panel, we plot the limits as a function of the mediator mass for
three representative DM masses, $10,~100~{\rm GeV}$ and $1~{\rm TeV}$.
In PandaX-II, the typical momentum transfer is $10\textup{--}50~{\rm MeV}$
for the DM mass $10~{\rm GeV}\textup{--}1~{\rm TeV}$ with a heavy mediator. Thus, for $m_\phi\gtrsim100~{\rm MeV}$,
the observed limits quickly approach to the results from the recent
WIMP searches in PandaX-II~\cite{Cui:2017nnn}. When the mediator
decreases, the signal spectrum peaks towards low recoil energies, as
shown in Figure~\ref{fig:s1}. Accordingly, the limits become weak, as
the detection efficiency decreases significantly for $S1$ below
$10~{\rm PE}$. For comparison, we also show the results converted from
Del Nobile \textit{et al.}~\cite{DelNobile:2015uua}, where the authors
reported limits on the mixing parameter in the context of an SIDM
model (see below) by recasting an early LUX result. We see that our
results improve significantly from previous ones. In the right panel,
we show the limits vs. the DM mass for $m_\phi=1,~10~{\rm MeV}$ and
$1~{\rm GeV}$. For $m_\phi=1~{\rm GeV}$, the exclusion limits agree
with our recent WIMP results~\cite{Cui:2017nnn}, but become
significantly weaker as $m_\phi$ decreases to 10 MeV or less.

\begin{figure*}[!htbp]
  \centering
  \includegraphics[width=0.45\linewidth]{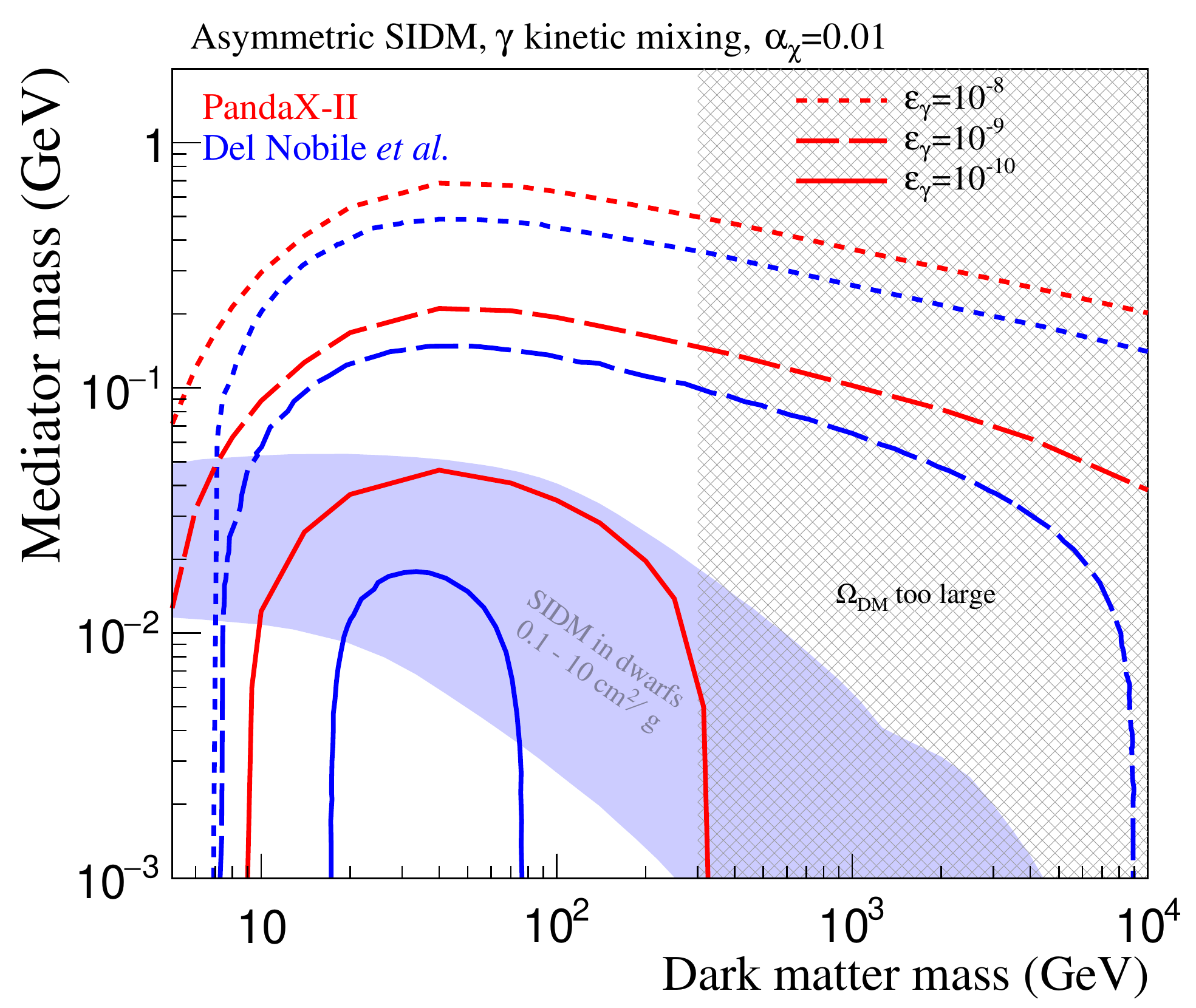}
  \includegraphics[width=0.45\linewidth]{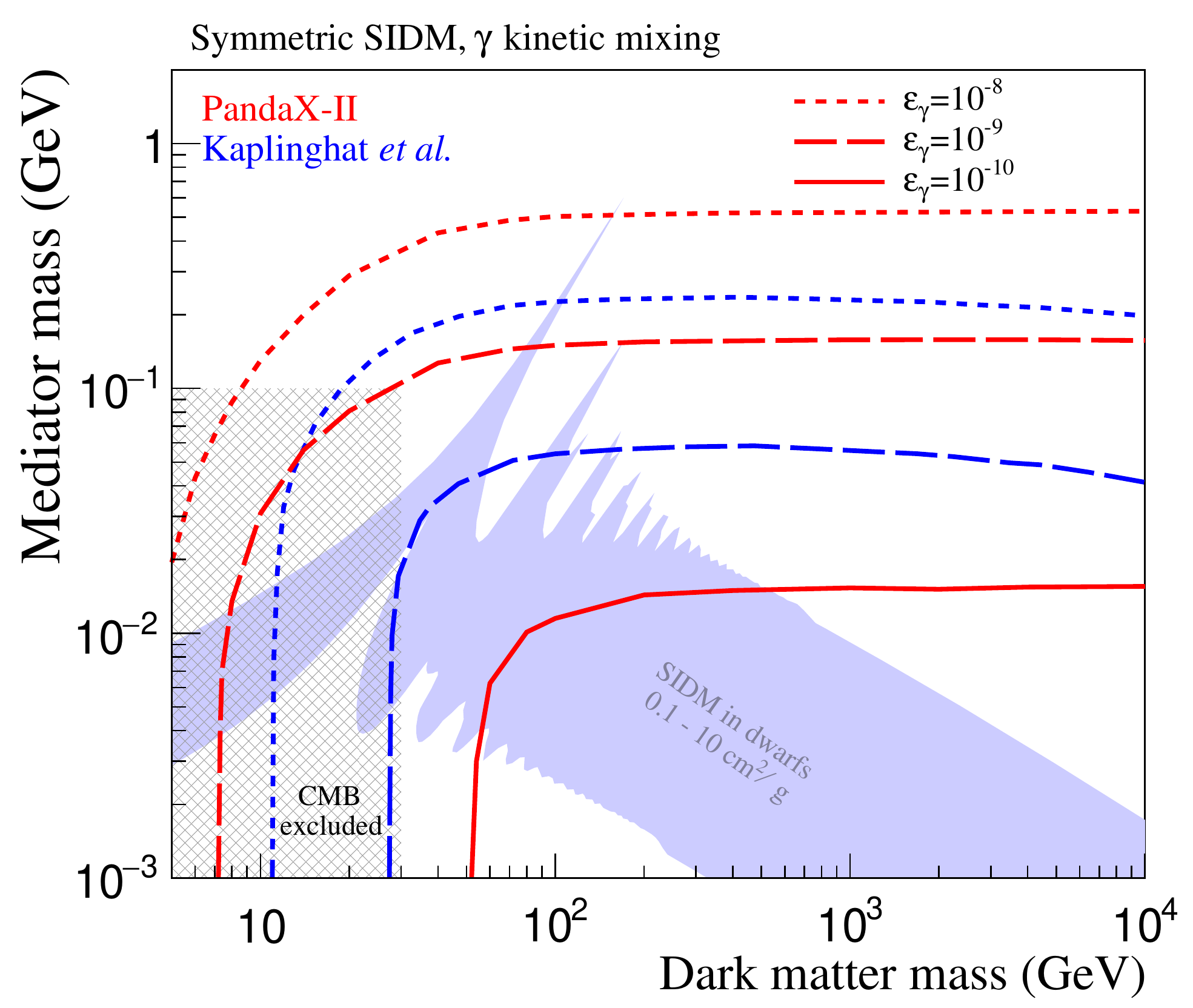}
  \includegraphics[width=0.45\linewidth]{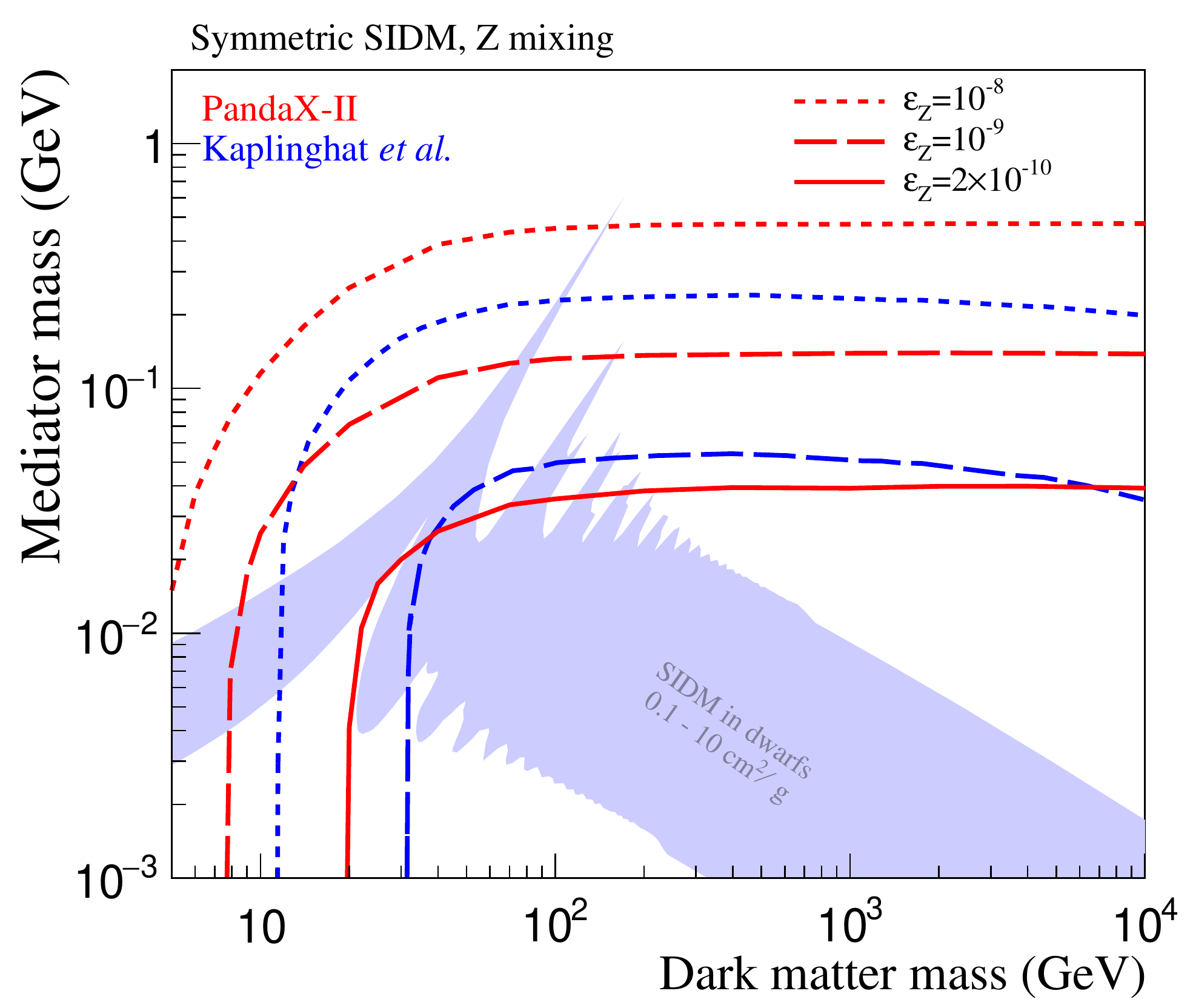}
  \includegraphics[width=0.45\linewidth]{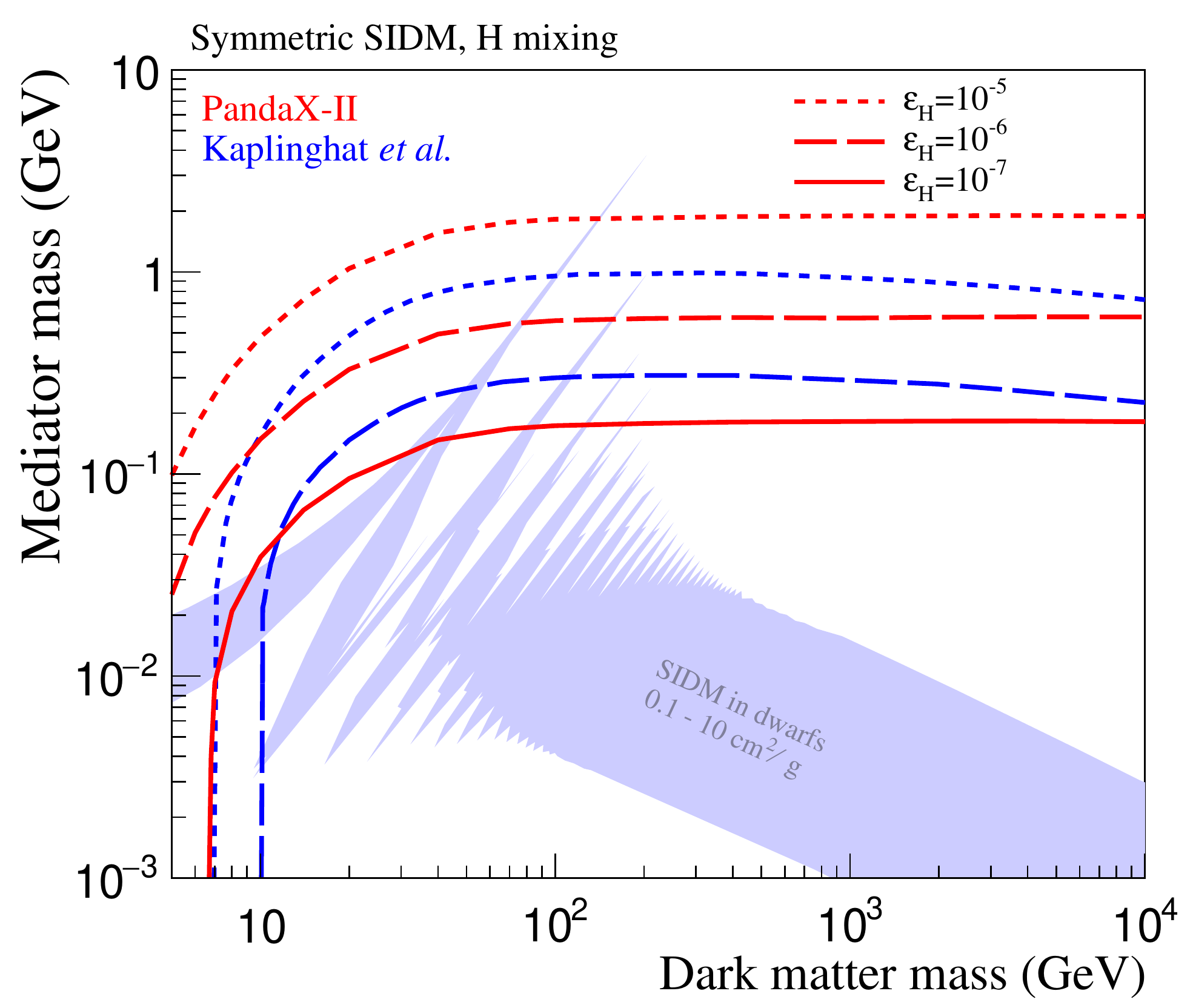}
  \caption{PandaX-II 90\% CL lower limits (red) on the
    ($m_{\phi},~m_{\chi}$) plane for four SIDM models: asymmetric SIDM
    with photon kinetic mixing (upper left), symmetric SIDM with
    photon kinetic mixing (upper right), symmetric SIDM with $Z$
    mixing (lower left) and symmetric SIDM with Higgs mixing (lower
    right). The shaded regions are favored by the observations in
    dwarf galaxies, taken
    from~\cite{Kaplinghat:2013yxa,DelNobile:2015uua}. Previous limits
    (blue) from~\cite{DelNobile:2015uua,Kaplinghat:2013yxa} are also
    shown for comparison. Cross shaded (upper): the DM abundance is
    too large for $m_\chi\gtrsim300~{\rm GeV}$ (left); the region for
    $m_\chi\gtrsim30~{\rm GeV}$ is excluded by the CMB observations
    (right), taken from~\cite{Kaplinghat:2013yxa}.}
  \label{fig:gammalimit2}
\end{figure*}

Our results have important implications in DM direct detection. For DM
models with a light mediator ($m_\phi\lesssim q$), the
contact-interaction approach can overestimate the direct detection
sensitivity by orders of magnitude, and a full treatment of the
scattering amplitude is required. From the model-building perspective,
light-mediator DM models provide a natural mechanism to suppress the detection signal by
populating events near the detector threshold limit. The analysis so far is based
on the general form given in Eq.~\ref{eq:general}. In what follows, we
apply our results to constrain the SIDM
models~\cite{Kaplinghat:2013yxa} by explicitly calculating
$\sigma|_{q^2=0}$. In these models, the force carrier mediating DM
self-interactions has a mass $\sim1~{\rm MeV}\textup{--}1~{\rm GeV}$.

Following~\cite{Kaplinghat:2013yxa}, we assume DM is a Dirac fermion
and it couples to a light mediator $\phi$. If $\phi$ is a vector
(scalar) particle, it can couple to SM fermions through $\gamma/Z$
(Higgs) mixing. The DM-nucleon cross section in the limit of
$q^2=0$ can be written as
\begin{equation}
  \sigma|_{q^2=0}= \frac{16\pi\alpha_{\mathrm{SM}}\alpha_{\chi}\mu^{2}_{p}}{m^{4}_{\phi}}\left[\frac{\epsilon_{p}Z+\epsilon_{n} (A-Z)}{A}\right]^{2},
  \label{sigmap}
\end{equation}
where $\alpha_{\mathrm{SM}}$ and $\alpha_\chi$ are the fine structure
constants in the visible and dark sectors, respectively,
$\epsilon_{p,n}$ are the effective proton or neutron couplings, and
$Z$ is the proton number of the nucleus. For photon kinetic mixing or
$Z$ mixing, $\epsilon_{p,n}$ are given by
\begin{eqnarray}
\label{eq:Zmixing}
  \epsilon_{p}=\epsilon_{\gamma} +
  \frac{\epsilon_{Z}}{4s_{W}c_{W}}(1-4s^{2}_{W}),~\epsilon_{n} =
  -\frac{\epsilon_{Z}}{4s_{W}c_{W}} \ ,
  \label{epsilon}
\end{eqnarray}
where $s_{W}$ and $c_{W}$ are the sine and cosine of the weak mixing
angle, and $\epsilon_{\gamma,Z}$ is the photon kinetic or $Z$ mixing
parameter. For Higgs mixing, they are
\begin{equation}
\label{eq:Hmixing}
\epsilon_{p,n}=\frac{m_{p,n}\epsilon_{H}}{e
  \mathrm{V}}(1-7f^{p,n}_{TG}/9),
\end{equation}
where $\epsilon_{H}$ is the Higgs mixing parameter, $e$ is the
electron charge, $\mathrm{V}$ is the vacuum expectation value of the
Higgs field, and $f^{p,n}_{TG}$ is determined by the gluon hadronic
matrix element, which we take
$f^{p,n}_{TG}=0.943$~\cite{Bhattacherjee:2013jca}.

We consider four cases. One is asymmetric SIDM with photon
kinetic mixing. Asymmetric SIDM arises from the possibility that DM
and anti-DM particles are not equally populated in the early universe
due to a primordial DM-number
asymmetry~\cite{Zurek:2013wia,Petraki:2013wwa}.  Other three are
symmetric SIDM with photon kinetic mixing, $Z$ mixing or Higgs
mixing. For asymmetric SIDM, we set $\alpha_\chi$ to be $0.01$, a choice
motivated by the value of the electromagnetic fine structure constant in the
SM~\cite{DelNobile:2015uua}. For symmetric SIDM, the DM relic density is set by the
annihilation process $\chi\bar{\chi}\rightarrow\phi\phi$, which sets
$\alpha_{\chi}$ values as~\cite{Kaplinghat:2013yxa},
$\alpha_{\chi}\approx4\times10^{-5}\times(m_{\chi}/\mathrm{GeV})$ for
photon kinetic mixing or $Z$ mixing, and
$\alpha_{\chi}\approx10^{-4}\times(m_{\chi}/\mathrm{GeV})$ for Higgs
mixing.

For each case, astrophysical observations set a preferred region in
the $m_{\phi}\textup{--}m_{\chi}$ plane, where the self-scattering
cross section per mass in dwarf galaxies is $\sim0.1\textup{--}10~{\rm
  cm^2/g}$~\cite{Kaplinghat:2013yxa,DelNobile:2015uua}. On the other
hand, for a given DM mass, direct detection experiments put a
constraint on the combination of the mixing parameter and the mediator
mass. To present our limits in the $m_{\phi}\textup{--}m_{\chi}$
plane, we will assume certain values of the mixing parameter. Note
that Kaplinghat \textit{et al.}~\cite{Kaplinghat:2013yxa} have
reinterpreted an early XENON100 WIMP search
result~\cite{Aprile:2012nq} to constrain the four cases, where a
constant momentum transfer was assumed in calculating the total signal
event. Furthermore, Del Nobile~\textit{et
  al.}~\cite{DelNobile:2015uua} simulated full energy spectra and
recasted results from early LUX~\cite{Akerib:2013tjd} and SuperCDMS
WIMP searches~\cite{Agnese:2014aze} to further constrain the
asymmetric one. The present study uses the largest data set to date,
and applies the complete analysis machinery in PandaX-II, including a
thorough modeling the detector response to signal and background based
on the calibration data.

Figure~\ref{fig:gammalimit2} shows the 90\% CL lower limits on the
($m_\phi,~m_\chi$) parameter region for four SIDM models. Our limits (red) are reported at three
$\epsilon_{\gamma}$ values, $10^{-8}$, $10^{-9}$, and $10^{-10}$.
Previous limits from Del Nobile \textit{et
  al.}~\cite{DelNobile:2015uua} by recasting a LUX result are included
for comparison. The SuperCDMS limits are significantly weaker and not shown here. The
shaded region is favored by observations in dwarf
galaxies. Overall, a heavier DM particle requires a lighter mediator
to enhance the self-scattering cross section, while keeping
$\sigma_{\chi\chi}/m_{\chi}$ constant. If $\epsilon_{\gamma}>10^{-9}$,
our results exclude all favored region with $m_{\chi}\gtrsim7~{\rm
  GeV}$. Even for $\epsilon_{\gamma}=10^{-10}$, we exclude a
significant part of the favored region for DM masses ranging from $10$
to $300~{\rm GeV}$. Previous limits are significantly weaker, in
particular for a small mixing parameter.

The remaining panels in Figure~\ref{fig:gammalimit2} are for symmetric
SIDM models, with photon kinetic mixing, $Z$ mixing or Higgs mixing.
Our limits are significantly stronger than the previous
ones~\cite{Kaplinghat:2013yxa}. The features in the shaded SIDM region
are due to the quantum resonant effect of attractive DM
self-interactions~\cite{Tulin:2012wi}. For photon kinetic mixing
$\epsilon_{\gamma}>10^{-9}$, our results exclude most of favored
region by observations in dwarf galaxies for $m_{\chi}\geq7~{\rm
  GeV}$. Even for $\epsilon_{\gamma}>10^{-10}$, we exclude a large
parameter space. Similarly, almost all favored heavy DM region is
excluded for $Z$ mixing and $\epsilon_{Z}>10^{-9}$. Our results are
not yet sensitive to $\epsilon_{Z}=10^{-10}$, but for
$\epsilon_{Z}=2\times10^{-10}$, we can exclude a large portion of the
favored region. For Higgs mixing, almost all favored region shown is
excluded if $\epsilon_{H}>10^{-6}$. We see that in most of the parameter region favored by the
astrophysical observations, our results are sensitive to the mixing parameter
as small as $\sim10^{-10}$ for photon kinetic or $Z$ mixing, and $10^{-7}$ for
Higgs mixing. The latter is weaker due to the suppression factor of $m_{p, n}/\mathrm{V}$ in Eq.~\ref{eq:Hmixing}.

In conclusion, using a combined data corresponding to a total exposure
of 54 ton day from the PandaX-II experiment, we have presented upper
limits on the DM-nucleon scattering cross section with the mediator
mass ranging from $1~{\rm MeV}$ to $1~{\rm GeV}$.  The
mediator mass plays a critical role in setting the exclusion limits
for these models and a full analysis of the scattering amplitude is
required. We further interpreted them to constrain the parameter space
in the context of the SIDM models with a light mediator mixing with SM
particles, complementing constraints from astrophysical
observations. These are the first kind of results reported by a direct
detection experimental collaboration. With more data from PandaX,
particularly the future multi-ton scale experiment at CJPL, we will
continue to probe the DM interaction with a light mediator and the
self-interacting nature of DM.

\begin{acknowledgments}
  This project has been supported by a 985-III grant from Shanghai
  Jiao Tong University, grants from National Science Foundation of
  China (Nos. 11435008, 11505112, 11525522 and 11755001), a grant from
  the Ministry of Science and Technology of China
  (No. 2016YFA0400301). We thank the support of grants from the Office
  of Science and Technology, Shanghai Municipal Government
  (Nos. 11DZ2260700, 16DZ2260200, and 18JC1410200), and the support from the Key
  Laboratory for Particle Physics, Astrophysics and Cosmology,
  Ministry of Education. This work is supported in part by the Chinese
  Academy of Sciences Center for Excellence in Particle Physics
  (CCEPP) and Hongwen Foundation in Hong Kong. Finally, we thank the
  following organizations for indispensable logistics and other
  supports: the CJPL administration and the Yalong River Hydropower
  Development Company Ltd. HBY acknowledges support from
  U. S. Department of Energy under Grant No. de-sc0008541.
\end{acknowledgments}

\bibliographystyle{apsrev4-1}
\bibliography{refs.bib}

\end{document}